\documentstyle[12pt,preprint,graphicx,natbib]{aastex}
\newcommand{\soho}{{\em SOHO{}}}
\newcommand{\pref}{\protect\ref}
\newcommand{\brr}{$\langle B_{||} \rangle_{r=20}$}

\begin{document}

\shorttitle{Energy Supply to the Solar Atmosphere}
\shortauthors{S.~W. McIntosh et~al.}
\title{Observations Supporting the Role of Magnetoconvection in Energy Supply to the Quiescent Solar Atmosphere}
\author{Scott W. McIntosh, Alisdair R. Davey, Donald M. Hassler}
\affil{Southwest Research Institute, Department of Space Studies,\\
1050 Walnut St, Suite 400, Boulder, CO 80302 USA}
\author{James D. Armstrong}
\affil{Institute for Astronomy, University of Hawaii, \\
4761 Lower Kula Road, Kula, HI 96790 USA}
\author{Werner Curdt, Klaus Wilhelm}
\affil{Max Planck Institute for Solar System Research,\\
 Max-Planck-Str. 2, 37191 Katlenburg-Lindau, Germany}
\author{Gang Lin}
\affil{Department of Electrical, Computer and Systems Engineering,\\
Rensselaer Polytechnic Institute, 110, 8th Street, Troy, NY 12180 USA}

\begin{abstract}
Identifying the two physical mechanisms behind the production and sustenance of the quiescent solar corona and solar wind poses two of the outstanding problems in solar physics today. We present analysis of spectroscopic observations from the {\em Solar and Heliospheric Observatory} that are consistent with a single physical mechanism being responsible for a significant portion of the heat supplied to the lower solar corona and the initial acceleration of the solar wind; the ubiquitous action of magnetoconvection-driven reprocessing and exchange reconnection of the Sun's magnetic field on the supergranular scale. We deduce that while the net magnetic flux on the scale of a supergranule controls the injection rate of mass and energy into the transition region plasma it is the global magnetic topology of the plasma that dictates whether the released ejecta provides thermal input to the quiet solar corona or becomes a tributary that feeds the solar wind.
\end{abstract}


\section{Introduction}

The relentless, convection-driven reprocessing of magnetic fields on the Sun has been dubbed the ``Magnetic Carpet'' \citep[e.g.,][]{Schrijver1997, Hagenaar1997, Title1998, Hagenaar1999}. Detailed observations of the Sun's photospheric magnetic field have been used to deduce that the turbulent motion of the ionized gas in the Solar interior causes small bipolar magnetic flux elements to be carried towards the photosphere. There, through their annihilation, they contribute the global magnetic structure of the outer solar atmosphere and are a source of energy to the corona and solar wind \citep[][]{Schrijver1998, PriestSchrijver1999, Priest2002}.

Numerical simulations of magnetoconvection \citep[][]{Cattaneo2003} and tectonic models of the magnetic carpet \citep[][]{Priest2002} show that the emerging bipolar elements are advected towards the boundaries of large convective cells with typical diameters of 15-35Mm \citep[e.g.,][]{Hagenaar1997}. These large supergranular convection cells are relatively easy to observe in the solar photosphere and chromosphere, and comprise the ``chromospheric network''. Over periods of many hours to days, the constant convective dredging of the magnetic field imposes a net magnetic polarity on a supergranule. This determines the polarity of the emerging magnetic dipoles which are annihilated through magnetic reconnection \citep[e.g.,][]{Parker1988, Parker1994, PriestForbes2000}. It has been postulated that energy stored in, and eventually released by, these small annihilated magnetic dipoles is the simplest and most efficient means of providing the 100 W/m$^{2}$ required to produce the Sun's ambient 2x$10^{6}$K corona \citep[][]{PriestSchrijver1999, Priest2002, Parker1994} and to load and accelerate mass into the solar wind \citep[][]{PriestForbes2000,Wang1998,Wang2004}. 

Recently, observations from the Solar Ultraviolet Measurement of Emitted Radiation \citep[SUMER;][]{Wilhelm+others1995}, the Extreme-ultraviolet Imaging Telescope \citep[EIT;][]{Boudine1995} and the Michelson Doppler Imager \citep[MDI;][]{Scherrer1995} instruments on the Solar and Heliospheric Observatory \citep[\soho;][]{Fleck+others1995} have provided insight into the origins of the fast solar wind in coronal holes \citep[][]{Hassler+others1999, Dammasch+others1999, Xia2003, Xia2004, Tu2005a, Tu2005b}. SUMER ``spectroheliogram'', or raster, observations were employed to quantitatively examine the origins of the solar wind by correlating Doppler velocity measurements of \ion{Ne}{8} \citep[formed at about 600,000K in the upper solar transition region;][]{Mazzotta1998} to a proxy for the gross super-granulation pattern of chromospheric network structure: bright \ion{Si}{2} emission (formed at about 10,000K). 

Results of these prior studies suggested a clear correlation between the observed \ion{Ne}{8} blue shift\footnote{In this paper we will use the convention that a blue shift is associated with a negative velocity and as such a blue Doppler shift indicates the presence of plasma moving towards the observer, possibly indicating an out-flow from the Sun.} and the chromospheric network pattern seen in \ion{Si}{2}. We show that while there is indeed a strong correlation between bright \ion{Si}{2} emission and \ion{Ne}{8} blue Doppler-shifts, it is limited to regions that occur at the intersections of three or more supergranules (see, e.g., Fig.~\pref{fig3}). We call these locations ``bright network vertices''.

We demonstrate that a substantial fraction of the blue shifted \ion{Ne}{8} plasma in coronal holes is not correlated with the bright \ion{Si}{2} emission that forms the supergranular boundaries or bright network vertices, consistent with the recent analysis of \citet[][]{Aiouaz2005}. We investigate the remaining portion of the blue shifted \ion{Ne}{8} plasma in the coronal hole to examine the mechanism behind the energetic processes taking place in the outer solar atmosphere.

We find that a critical - and largely overlooked - step towards understanding the excess \ion{Ne}{8} Doppler shift lies in the interpretation of the \ion{C}{4} emission line, formed in the transition region at roughly 100,000K between \ion{Si}{2} and \ion{Ne}{8}. Data available at this thermodynamic mid-point allows us to physically couple the spatial variation of the lower temperature \ion{Si}{2} and higher temperature \ion{Ne}{8} plasmas to form an entirely new interpretation of these well studied data.

We present the self-consistent analysis, comparison, and interpretation of the results derived from spectroscopic observations of coronal hole and quiet Sun regions which point to the magnetic carpet's constant recycling of magnetic flux as a significant energy supply mechanism to the outer solar atmosphere. Indeed, we deduce that the difference in the form of the energy provided to the plasma stems, rather simply, from whether or not its magnetic topology is ``open'' (coronal hole; largely kinetic) or ``closed'' (quiet Sun; largely thermal) to interplanetary space.

\section{Observations and Data Reduction}\label{sec:data}
Many investigations into SUMER raster observations of coronal holes have focused on the Sun's polar regions \citep[][]{Hassler+others1999, Dammasch+others1999, Xia2003, Tu2005a}. We instead concentrate on an equatorial coronal hole (hereafter referred to as a coronal hole) to minimize interpretative error and any line-of-sight effects on the magnetic fields and Doppler images. Viewing an equatorial region largely removes the need to perform latitude-dependent, transforms to the data. Indeed, such transforms to the Doppler data are not simple and require a significant amount of {\it a priori} information of the magnetic structures present because the plasma and its observed flows are intimately bound to those structures.

We use two SUMER rasters in the 1530-1555\AA{} spectral range. This portion of the Solar UV spectrum contains emission lines of singly ionized Silicon (\ion{Si}{2}) with a rest wavelength of 1533.43\AA{}, seven times ionized Neon (\ion{Ne}{8}) at 1540.84\AA{} (in second spectral order) and two lines of three times ionized Carbon (\ion{C}{4}) at 1548.20\AA{} and 1550.77\AA{}. The two \ion{C}{4} lines behave identically and so we only consider information from the former in this paper as it has better signal-to-noise. These three emission lines span the highly dynamic solar Transition Region; the thermodynamic region of the solar atmosphere coupling the chromosphere at 10,000K to the upper transition region or low solar corona at nearly 1x$10^{6}$K.

\begin{figure}
\centering
\epsscale{0.75}
\plotone{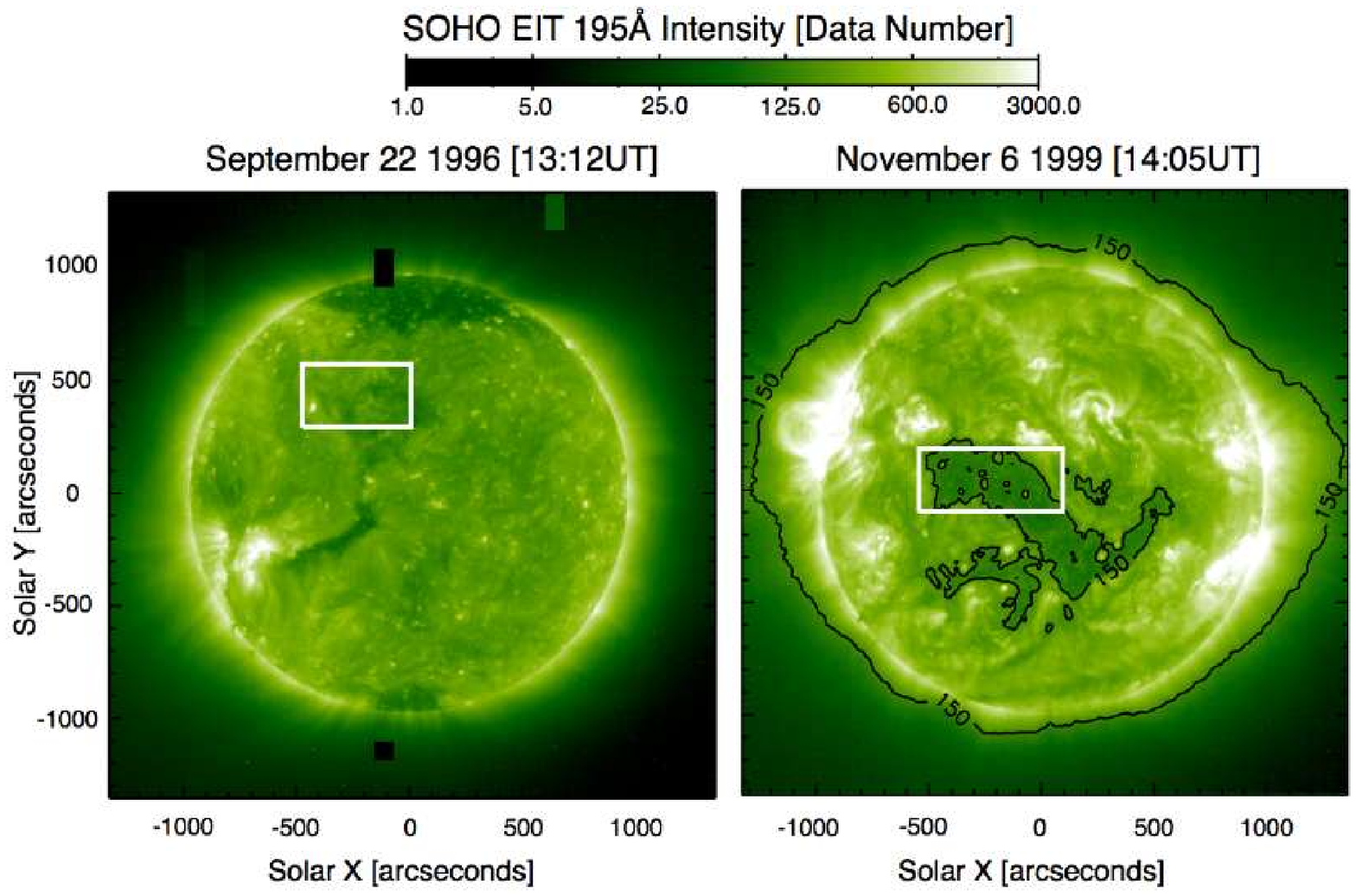}
\caption{\soho{} Extreme-Ultraviolet Imaging Telescope (EIT) images of the Sun's Corona (seen in 1.5 MK, \ion{Fe}{12} 195\AA) on 1996 September 22 (left panel) and November 6, 1999 (right panel). The right panel shows the outline of the equatorial coronal hole region studied in this paper as a black closed contour (at the 150 Data Number level). Also shown are the regions of the solar atmosphere concurrently mapped by the SUMER instrument (white outlined rectangles) on both days. \label{fig1}}
\end{figure}

Each SUMER raster that we have studied uses Detector A, the 1\arcsec x 300\arcsec{} slit (SUMER Slit 2), exposure times of 150s, and a 3\arcsec{} raster step size from West to East in direction. The first raster is of an coronal hole, from 1999 November 6 (14:00-00:00UT), and the second is of a well-studied \citep[][]{Hassler+others1999, Dammasch+others1999, Xia2004, Tu2005b} portion of quiet Sun from 1996 September 22  (00:40-08:28UT). The EIT 195\AA{} reference images (presented in Fig.~\pref{fig1}) provide contextual data on the hot solar corona \citep[emission of eleven times ionized iron, \ion{Fe}{12}, at about 1.5x10$^{6}$K; ][]{Boudine1995} nearest to the start of each observation along with the rectangular region rastered by SUMER. Similarly, we employ the full-disk photospheric MDI line-of-sight magnetograms nearest to the start of each observation to explore the magnetic environment in SUMER's field of view.

The robust absolute wavelength calibration of the observed line spectra in SUMER raster observations is critical when attempting to accurately measure and isolate the source regions of the fast solar wind. \citet{Davey2006} quantified the small (persistent) electronic imperfections of SUMER detector A and developed an improved method to correct for the systematic shifting of the resulting Solar UV spectra when acquired in raster mode. While the detector imperfections are small in pixel terms ($\pm$0.5 pixels), in terms of velocity, at 1535\AA, these translate to $\pm$4km/s \citep[][]{Wilhelm+others1995}. At this magnitude, they influence any attempt at absolute physical interpretation of the data.

By mapping the locations of ten laboratory measured emission lines of neutral silicon (\ion{Si}{1}) in the 1530-1555\AA{} range at each spatial row of pixels, \citet{Davey2006} performed a self-consistent wavelength calibration for nine SUMER detector A raster observations, including the two discussed above. The result of this calibration is a set of Doppler velocities in the principal science lines (\ion{Si}{2}, \ion{C}{4} and \ion{Ne}{8}) to an accuracy of about $\pm$1km/s. The corrected data form the ``cleanest'' set of SUMER raster observations available, and their contents have prompted the investigation presented here.

We note that, for the present analysis, following the instrumental discussion of Davey et~al. (2006; Sect.~4) and the topological argument in McIntosh et~al. (2006; Sect.~3) we use 770.420\AA{} as the rest wavelength of the \ion{Ne}{8} emission line to compute the Doppler velocity maps presented in this Paper. This choice agrees with one of the laboratory determinations of the rest wavelength for this emission line \citep[][]{Fawcett1961} and will be the subject of a future publication (Davey \& McIntosh 2006 \-- in preparation).

\subsection{Supergranular Boundaries}
\begin{figure}
\centering
\epsscale{0.75}
\plotone{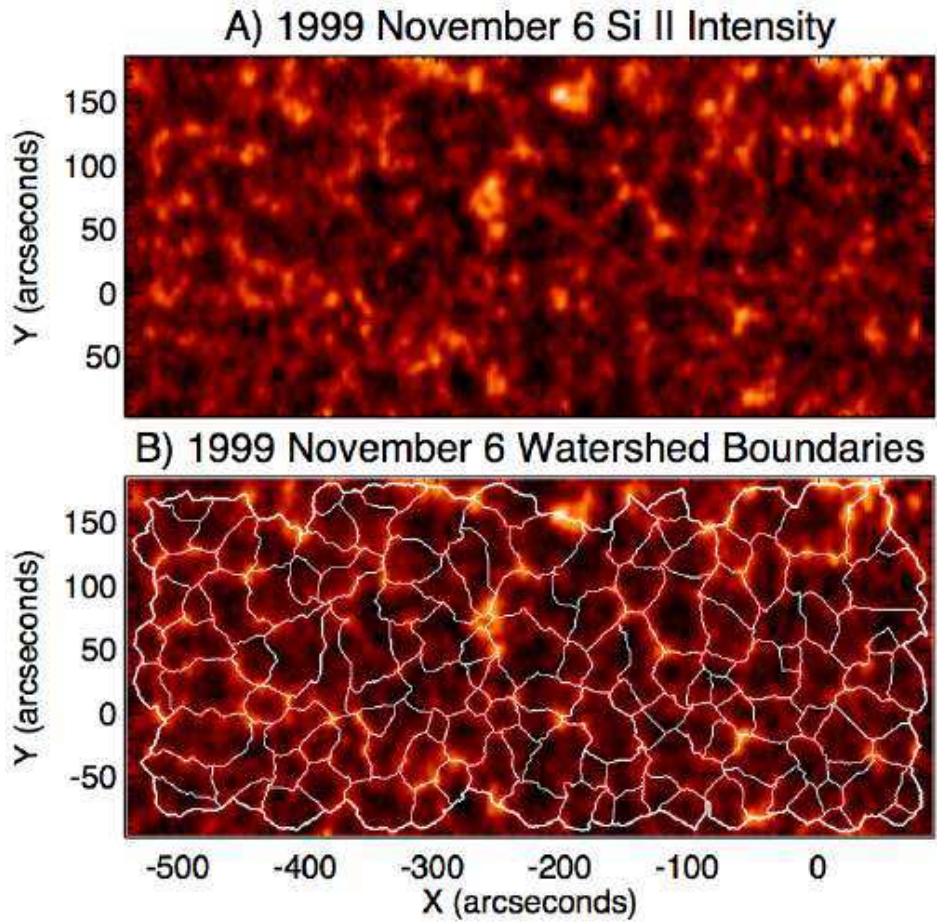}
\caption{Comparison of the 1999 November 11 \ion{Si}{2} intensity raster and the numerically-determined watershed segmentation boundaries that we use throughout as a proxy for the underlying supergranular network cell pattern. Note that for the bulk of the statistics presented later, we use a version of the boundaries that is artificially thickened to three pixels (adding one pixel to either side of the boundary), giving them an assumed (artificial) thickness of 9\arcsec. \label{fig2}}
\end{figure}

In Fig.~\pref{fig2}, we show the empirically determined supergranular boundaries derived from the \ion{Si}{2} intensity image (panels A), using an enhanced watershed segmentation technique \citep[][]{Lin2003}. Watershed segmentation uses the derivative of the intensity image as a topographic map to determine numerical ``watershed basins''. The places where the watershed basins meet form an initial watershed boundary to which a model-based object-merging is used to eliminate over-segmentation of the \ion{Si}{2} image \citep[][]{Lin2005}. These numerically determined supergranular boundaries can be readily compared to the intensity pattern observed (panel A) as well as the heuristic boundaries that have previously been published for the 1996 September 22 SUMER observations \citep[cf., Fig.~\pref{fig3} of][]{Hassler+others1999}.

In the large majority of the analysis to follow, we will need to artificially thicken the computed watershed boundaries in order to produce distributions of quantities belonging to the supergranular cell boundaries and interiors. The thickening is simply achieved by allowing the boundary mask (white contour) to expand by one pixel on either side. This effectively forces the boundary thickness to be 3 pixels (9\arcsec) which appears to be consistent with those observed. An analysis of the derived supergranular network boundaries and their expansion above the photosphere is provided by \citet{Aiouaz2006}.

\section{Data Analysis}\label{sec:discuss}
In this section will use Figs.~\pref{fig3} through~\pref{fig11} to investigate the multithermal structure of the solar transition region, its relation to the supergranular network structure and magnetic environment threading the plasma using a single set of high quality SUMER observations. To clarify the details of the analysis we break this section of the paper into three parts, the first discusses the comparison of the spectroscopic data [\ion{Si}{2}, \ion{C}{4} and \ion{Ne}{8}] from the 1996 September 22 quiet Sun and 1999 November 6 coronal hole rasters, the second places the spectroscopic in context with the photospheric magnetic field and simple diagnostics derived from it while the third combines and compares the key factors of the analysis of the spectroscopic and magnetic diagnostics in the two plasma regimes. 

\subsection{Spectroscopic Data: Quiet Sun Vs. Coronal Hole}
\begin{figure}
\centering
\plotone{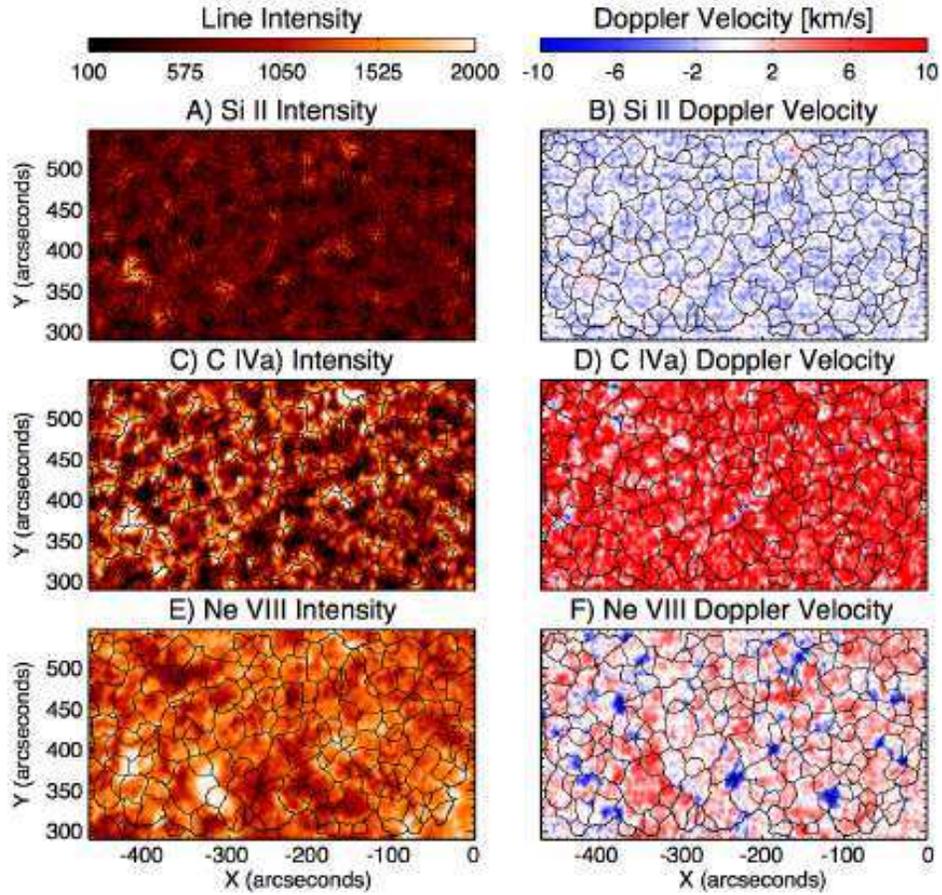}
\caption{The SUMER raster images of the quiet Sun region observed on 1996 September 22 (intensity images on the left with the corresponding line Doppler shift on the right). In each of the panels we show the inferred supergranular cell boundaries that are determined from the \ion{Si}{2} Intensity image as thin black lines.\label{fig3}}
\end{figure}

\begin{figure}
\centering
\plotone{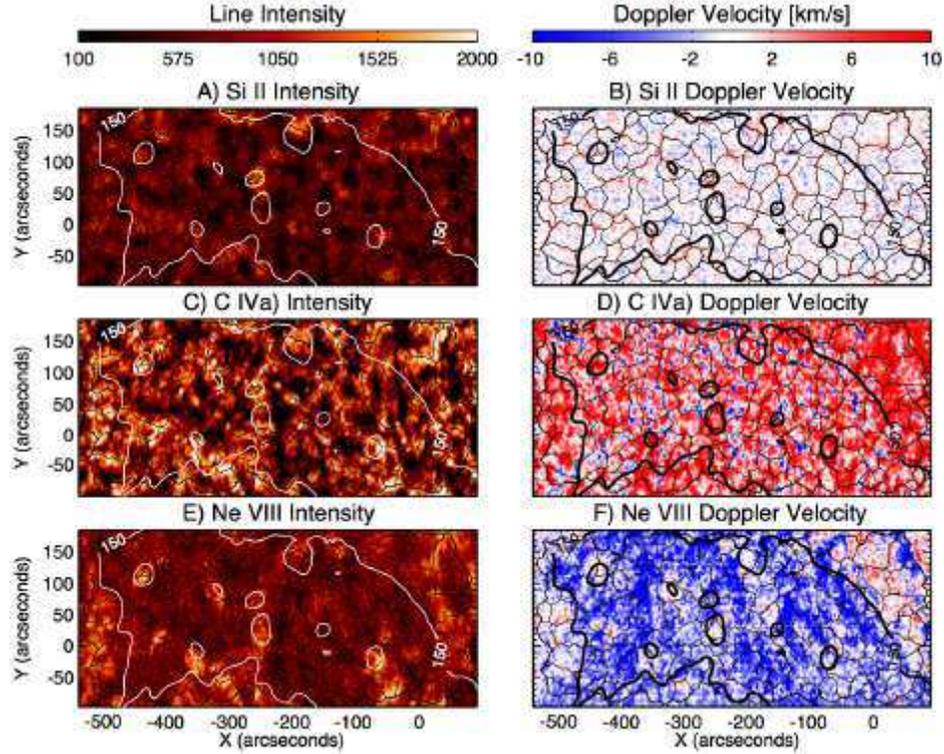}
\caption{The SUMER raster images of the equatorial coronal hole region observed on 1999 November 6 (intensity images on the left with the corresponding line Doppler shift on the right). In each of the lower panels, we show the EIT coronal hole boundary and the inferred supergranular cell boundaries determined from the \ion{Si}{2} Intensity image as thin black lines. On each panel, we show the coronal hole boundary as inferred from the EIT \ion{Fe}{12} emission (at a level of 150 Data Numbers) from the EIT \ion{Fe}{12} 195\AA{} image taken closest to the start of the SUMER raster scan. \label{fig4}}
\end{figure}

\begin{figure}
\centering
\plotone{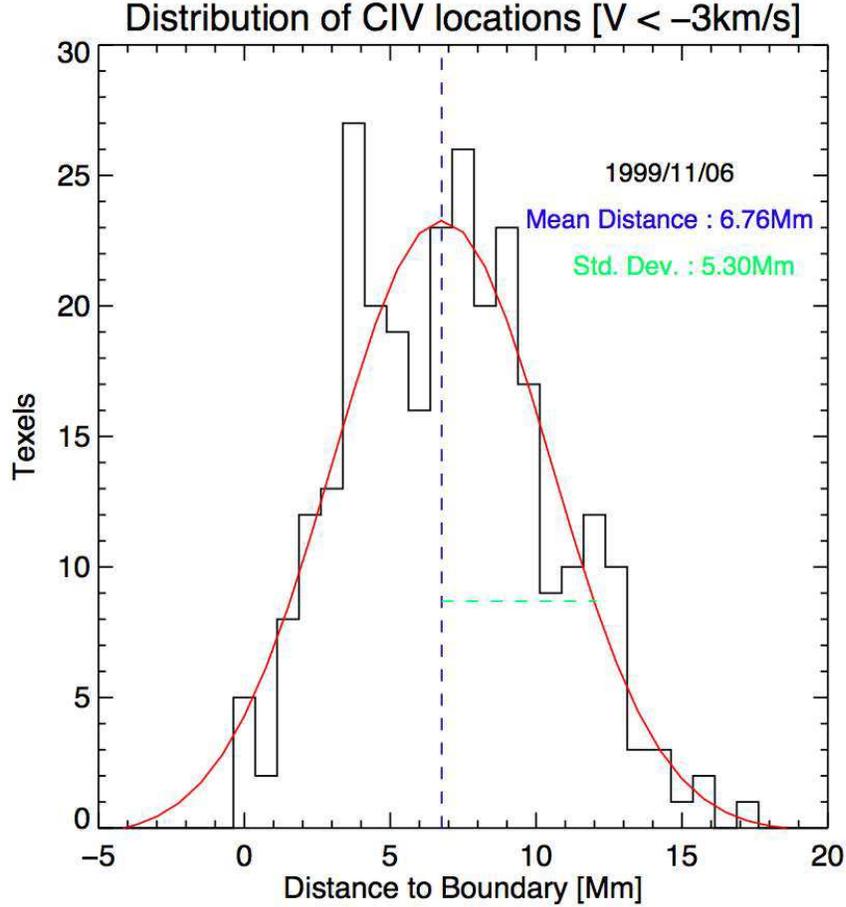}
\caption{The distance distribution of the larger \ion{C}{4} blue Doppler shifts ($<$ -3 km/s) compared to the  watershed segmentation supergranular boundary for the equatorial coronal hole raster of 1999 November 6. Using a method that has previously been employed to identify and locate Extreme Ultraviolet ``bright points'' in EIT images \citep[][]{McIntosh2005}, we locate regions of large blue shift and measure their distance from the inferred supergranular boundary. We see that only 8\% of the strong \ion{C}{4} blue shifts lie within 2Mm of the boundary. They have a mean distance of 6.76Mm. \label{fig5}}
\end{figure}

Figures~\pref{fig3} and~\pref{fig4} show the \ion{Si}{2}, \ion{C}{4} and \ion{Ne}{8} intensity and Doppler velocity rasters of the quiet Sun and coronal hole rasters of 1996 September 22 and 1999 November 6 respectively. The panels of each figure show the watershed boundaries computed from the \ion{Si}{2} intensity. Comparing panel B of both figures, we see that the \ion{Si}{2} Doppler velocities on the boundaries are predominantly red shifted by about 2km/s, while the -1km/s blue shifts in the interiors may represent convective ``overturn'' flow although this measurement is comparable to the $\pm$1km/s uncertainty in the Doppler velocity measurements \citep[][]{Davey2006}.

The \ion{C}{4} emission in panel C of each figure forms an apparently high intensity contrast proxy for the chromospheric network \citep[also noted by][]{Aiouaz2005} while, for the same panel of Fig.~\pref{fig4}, the emission inside the coronal hole appears weaker and the cell interiors are better defined (apparently darker, even in the logarithmic scale of the image). From panel D of Fig.~\pref{fig3} we see that there are very few locations in the field of view where the \ion{C}{4} emission is blue shifted (4\% of the total number of pixels in the image have blue shifts greater than 3km/s) and these locations appear to surround the bright network vertices. However, the coronal hole \ion{C}{4} Doppler velocity image (panel D of Fig.~\pref{fig4}) there is an apparently larger amount of blue-shifted \ion{C}{4} plasma inside the coronal hole boundary (21\% of the number of image pixels inside the coronal hole boundary contour) but the locations appear to show no significant correlation with the network, except that very few exist on the derived supergranular boundaries. This is remarkably close to previous measurements from sounding rocket flights by \citet{Dere1989} who demonstrated that 26\% of the observed coronal hole showed \ion{C}{4} blue shifts while only 7\% of the quiet Sun region did. In fact, using the distribution of \ion{C}{4}\--boundary distances shown in Fig.~\pref{fig5}, we can show that the large \ion{C}{4} blue shift regions have a mean distance of 6.7Mm from the supergranular boundary and only 8\% lie within 2Mm of the boundary. 


The \ion{Ne}{8} intensity in panel E of Fig.~\pref{fig3} is almost uniformly bright and appears to have lost the strongly contrasting emission from the chromospheric network. The same panel of Fig.~\pref{fig4} shows a marked decrease of the \ion{Ne}{8} intensity in the coronal hole. Clearly the most dramatic change in the spectroscopic data between the quiet sun and coronal hole lies in the amount and locations of the blue shifted \ion{Ne}{8} plasma (compare panel F of both figures). While the quiet Sun shows blue-shifts regions that overlie bright supergranular vertices \citep[at the junctions of three or more supergranules that radiate strongly in \ion{Si}{2}, e.g.,][]{Hassler+others1999,Xia2004,Tu2005b} there is a large portion of the coronal hole plasma that is blue shifted. The coronal hole \ion{Ne}{8} blue shifts are not solely located at the bright supergranular vertices and there appears to be little or no clearly visible tie to the supergranular network pattern (contrary to the result of \cite{Hassler+others1999}, but consistent with the recent analysis by \cite{Aiouaz2005} on the same SUMER raster data). It is the profound and apparently (physically) coupled differences between the emission and Doppler patterning of the quiet sun and coronal hole in the 100,000K (\ion{C}{4}) and 600,000K (\ion{Ne}{8}) transition region plasma that have provoked this investigation. 

\begin{figure}
\centering
\plotone{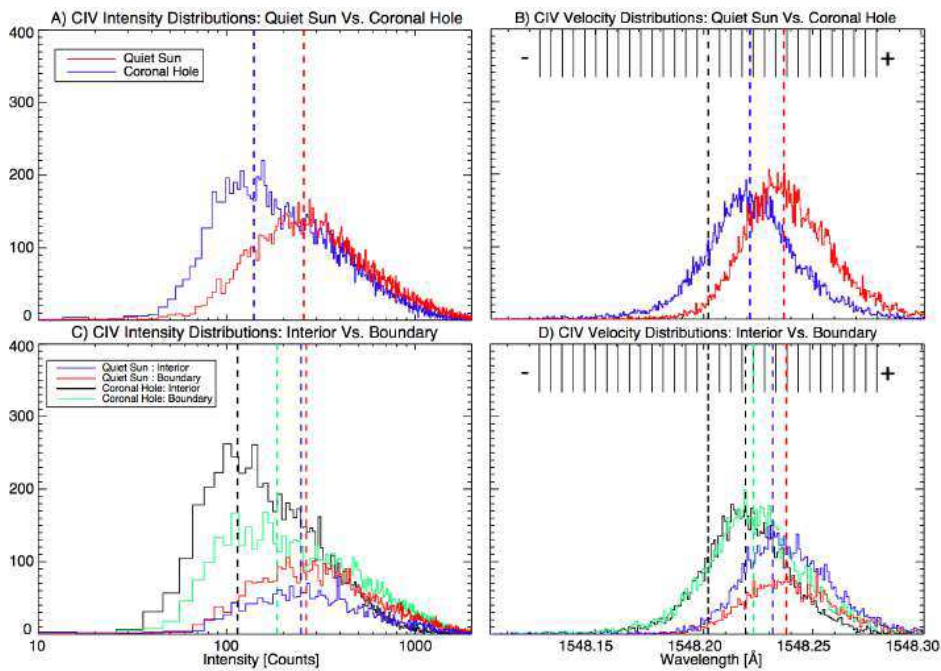}
\caption{Distributions of the \ion{C}{4} intensities (left column) and Doppler velocities (right column) for the data presented in Figs.~\pref{fig3} and~\pref{fig4}. While the top row of pixels show the distributions partitioned for the entire coronal hole and quiet Sun regions, the bottom row shows the breakdown of these into the supergranular cell boundary and interior components. In each panel, the vertical dashed lines mark the position of the distribution mean of the same color and thickness. In panels B and D, the thick black (dashed) line indicates the rest wavelength of the emission line, and the vertical ticks mark off one km/s increments in the Doppler velocity of the line. Compare with Fig.~11 of \citet{Davey2006}. \label{fig6}}
\end{figure}

\begin{figure}
\centering
\plotone{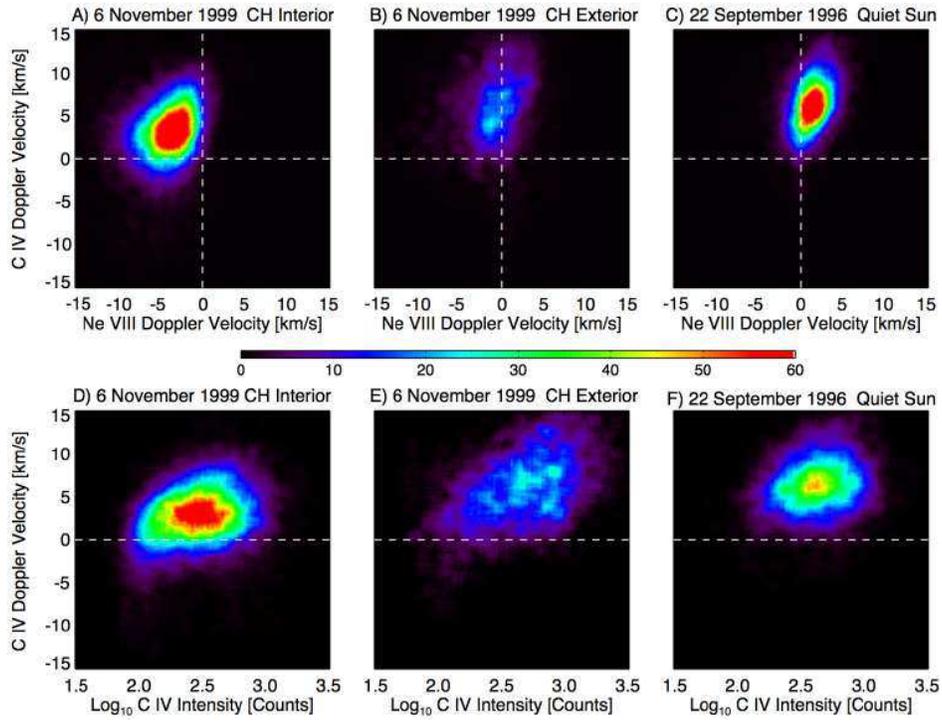}
\caption{Scatter diagrams to quantify the correspondence of \ion{C}{4} Doppler velocities with those of \ion{Ne}{8} (top row) and the logarithm of \ion{C}{4} intensity (bottom row) for supergranular interiors. From left to right the columns correspond to quantities that are in the computed supergranular network cell interiors inside the 1999 November 6 coronal hole (left), outside the coronal hole (center) and for the 1996 September 22 quiet Sun raster (right). The last two columns should behave identically. \label{fig7}}
\end{figure}

Figure~\pref{fig6} shows the distributions of the \ion{C}{4} spectroscopic diagnostics in each plasma regime. Panels A and B show the intensity and Doppler velocity distributions in the quiet Sun (red histogram) and coronal hole (blue histogram) data. Panels C and D show how these distributions break down into their supergranular network and supergranular interior components based on the watershed segmentation masks (see figure legend for histogram labeling). From panels A and B, we see that the \ion{C}{4} emission inside the coronal hole is reduced by almost 30\% (blue histogram) compared to that of the quiet Sun (red histogram) while the corresponding Doppler velocities change from $\sim$4km/s to $\sim$7km/s. The \ion{C}{4} emission contrast is most obvious in the cell interiors of the coronal hole (panel C) where the emitted intensity drops by a further 20\% from the supergranular boundary value. The contrast change between quiet Sun supergranular boundaries is not as profound (on the logarithmic scale used) but is very apparent in Figs.~\pref{fig3} and~\pref{fig4}. Further, we see that the \ion{C}{4} Doppler shifts on the supergranular boundaries are typically stronger (by $\sim$1km/s) than those in the supergranular interiors in both cases. These points are consistent with previous comparisons \citep[][]{Gebbie1981, Dere1989, Warren+others1997, Wilhelm2002}. We note that the presence of the large pervasive red shifts on the supergranular boundaries (of the quiet sun and coronal hole regions alike) is indicative of sub-spatial resolution magnetic fields advected there by the convective flow-field \citep[][]{Berger2001}, granular flow driven reconnection \citep[][]{Hansteen1996} or downward propagating acoustic waves \citep[][]{Hansteen1993}.

Figure~\pref{fig7} shows a set of scatter diagrams relating the \ion{C}{4} Doppler velocities with their \ion{Ne}{8} counterparts (top row of panels) and the \ion{C}{4} intensities (bottom row of panels) in the supergranular cell interiors of the coronal hole interior, exterior and quiet Sun\footnote{We make a distinction here with coronal hole exterior and interior of the 1999 November 6 data to demonstrate that (statistically at least) the coronal hole exterior (the region outside of the SOHO/EIT 150DN contour) behaves identically to the quiet Sun data of 1996 September 22, as we would expect.}. From panel A we see a strong correspondence between the coronal hole \ion{C}{4} and \ion{Ne}{8} blue shift locations. Close inspection of the panel (and the spectroscopic images) shows that a very large fraction (92\%) of the large \ion{C}{4} coronal hole blue shift locations underlie \ion{Ne}{8} blue shifts of the same or higher magnitude. This, we feel, indicates a direct physical connection between the two. Conversely, in the quiet Sun, we see that very few of the pixels with \ion{C}{4} blue shifts underlie \ion{Ne}{8} blue shifts; as we have stated above the large quiet Sun \ion{C}{4} blue shifts neighbor the large supergranular vertex \ion{Ne}{8} blue shifts instead, a point that we substantiate in the following subsection. In the coronal hole exterior (B) and quiet Sun (C), we see that the mean \ion{C}{4} supergranular cell interior Doppler shift has indeed increased by about 3km/s \citep[consistent with][and Fig.~\pref{fig6}]{Gebbie1981, Dere1989, Wilhelm2002}, with corresponding mean \ion{Ne}{8} velocities in the coronal hole exterior of $\sim$0km/s and quiet Sun of $\sim$1km/s, respectively. Note that there are very few blue shifted pixels in either of the latter regions. From panel D we see that the strongest blue shifted \ion{C}{4} pixels occur in the darkest portions of the coronal hole - noting again that the mean coronal hole supergranular interior \ion{C}{4} intensities are reduced by about 40\% compared to their counterparts in panels E and F (cf. Fig.~\pref{fig6}). \citet{Dere1989} noted no distinct correlation of the \ion{C}{4} blue shifts with bright intensity features. Since we see that the \ion{C}{4} blue shifts in the coronal hole exist predominantly in the (dark) supergranular interiors we can only speculate that a lack of spatial correspondence to anything bright made them appear to be insignificant.

\subsection{Spectroscopic Relationship to The Magnetic Topology}
Clearly, the relationship between the Doppler shift and intensity patterning of \ion{C}{4} and \ion{Ne}{8} in both regions is complex, but their behavior appears to be strongly coupled. We now investigate how the magnetic field observed in each region can add further evidence to help us explore the magnetic connection between the cool (\ion{Si}{2}) and hot (\ion{Ne}{8}) plasmas. This allows us to compare our results with previous investigations of the magnetic environments in coronal holes and quiet Sun \citep[][]{Hassler+others1999, Dammasch+others1999, Xia2003, Tu2005a, Aiouaz2005}.

\begin{figure}
\centering
\plotone{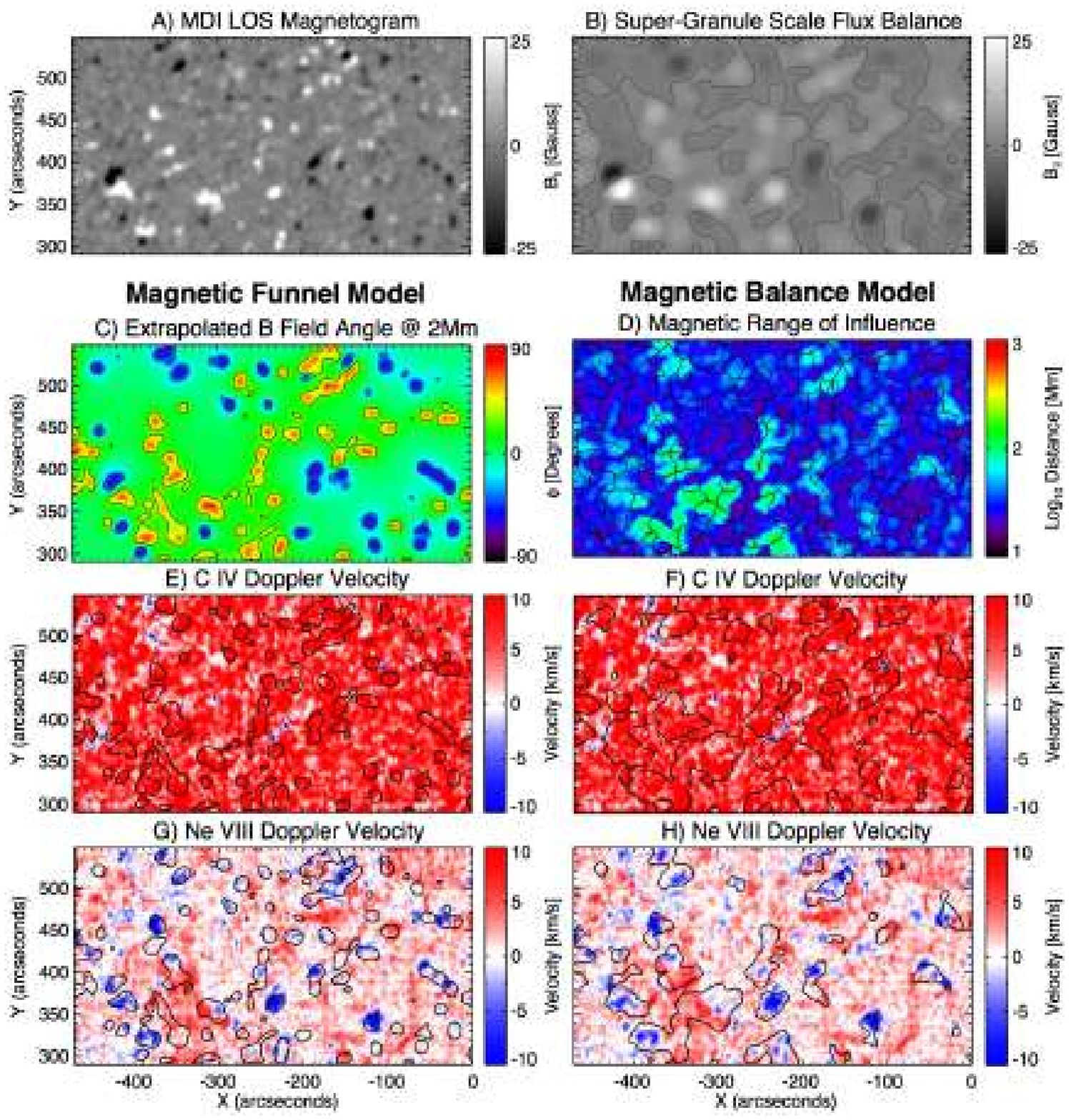}
\caption{Comparison of two simple magnetic models and their correlation to the \ion{C}{4} and \ion{Ne}{8} Doppler velocity maps from the quiet Sun SUMER raster of 1996 September 22. We show the MDI line-of-sight magnetogram (A) and the 20Mm smoothed MDI line-of-sight magnetogram \citep[\brr;][]{McIntosh2006}. In panel C we show the extrapolated magnetic field angle (to the vertical) at a height of 2Mm. We identify regions where the absolute value of the field angle is greater than or equal to 35 degrees as belonging to a magnetic funnel with sold black contours. These same regions are then compared to the \ion{C}{4} and \ion{Ne}{8} Doppler velocity maps (panels E \& G). In panel D, we show the ``Magnetic Range of Influence'' \citep[MRoI;][]{McIntosh2006} map derived from the magnetogram in panel A along with the watershed boundaries to demonstrate the net magnetic flux on any particular cell. We isolate regions where the MRoI is greater than 50Mm with solid black contours in panels F \& H. \label{fig8}}
\end{figure}

\begin{figure}
\centering
\plotone{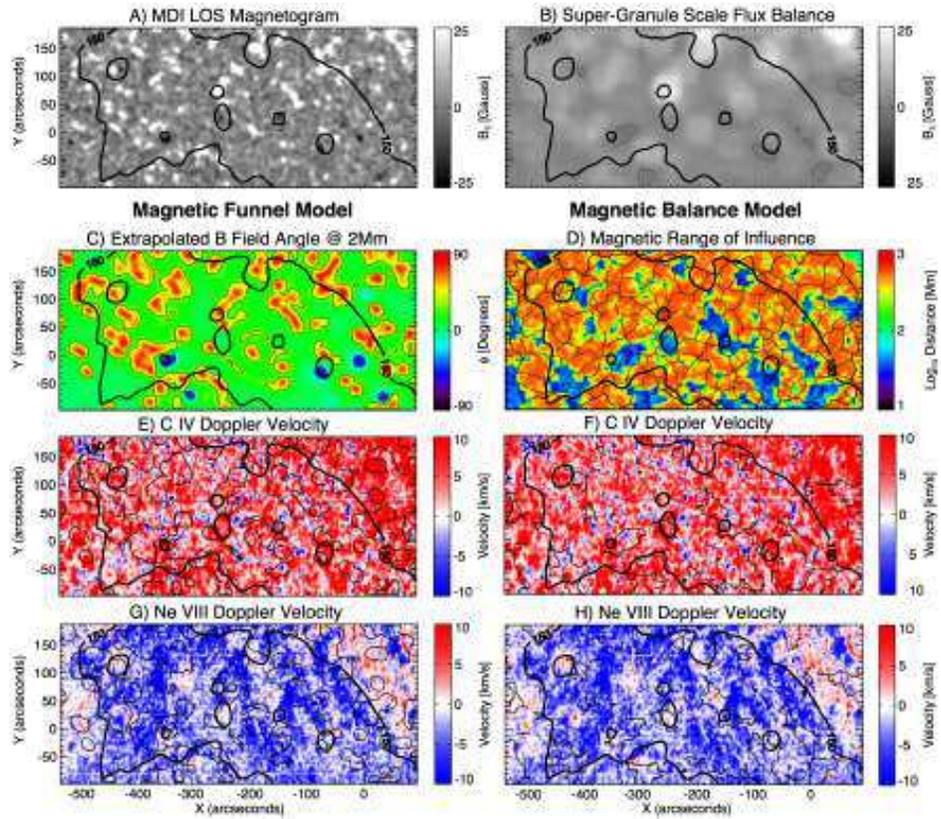}
\caption{As for Fig.~\pref{fig8} except that we now compare the simple magnetic models with the \ion{C}{4} and \ion{Ne}{8} Doppler velocity maps from the SUMER equatorial coronal hole raster of 1999 November 6. Again, we illustrate the coronal hole boundary by drawing the 150 Data Number contour in the EIT \ion{Fe}{12} 195\AA{} emission. \label{fig9}}
\end{figure}

\begin{figure}
\centering
\plotone{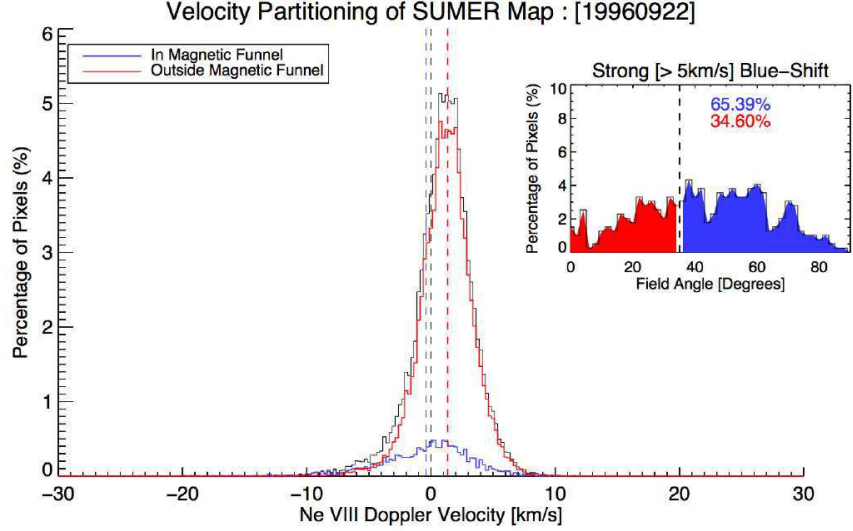}
\caption{Velocity partitioning plot for the quiet Sun SUMER raster of 1996 September 22. In the main body of the figure, we show the distribution of the pixels in the \ion{Ne}{8} Doppler shift map along with those which are inside magnetic funnels (blue) and those that are  outside (red). The inset plot shows the distribution of large ($<$ -5km/s) \ion{Ne}{8} blue shifts by field angle. We see that $\sim$65\% of the pixels with large blue shifts occur in funnels and, as such, back up the model of \citep[][]{Xia2003,Tu2005b} in the quiet Sun. \label{fig10}}
\end{figure}

\begin{figure}
\centering
\plotone{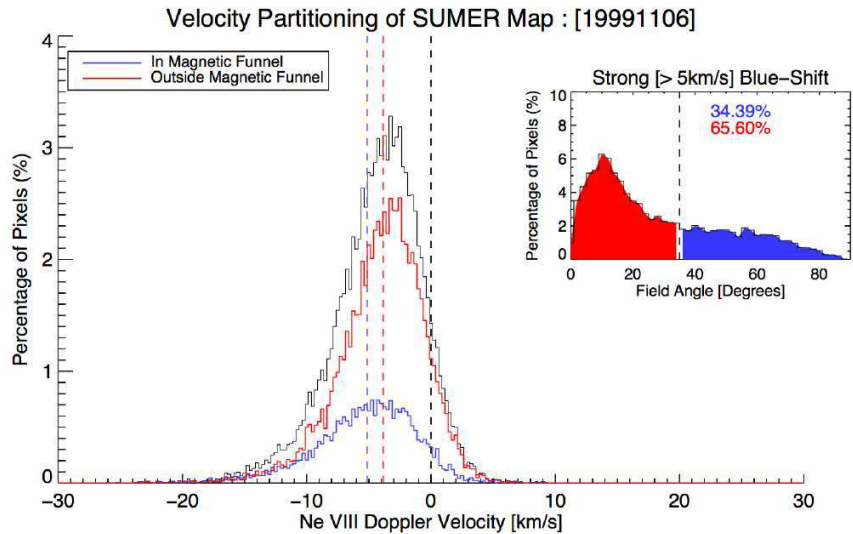}
\caption{As for Fig.~\pref{fig10}. Now we perform the same test for the equatorial coronal hole region of 1999 November 6 and can see that by far the largest portion (65\%) of strong \ion{Ne}{8} blue shifts do not occur magnetic funnels. \label{fig11}}
\end{figure}

Figures~\pref{fig8} and~\pref{fig9} show the comparison of two different models of the magnetic environment with the Doppler velocity maps of \ion{C}{4} and \ion{Ne}{8} in the quiet Sun and coronal hole, respectively. Panels A and B show two representations of the underlying photospheric magnetic field at the start of the SUMER raster observation: (A) the raw MDI magnetogram and (B) the same magnetogram smoothed by a circular filter of radius of 20Mm \citep[\brr;][]{McIntosh2006}. The latter shows the presence of net magnetic field polarities at a spatial scale commensurate with a supergranule. The fine black line in panel B shows the magnetic neutral line where positive and negative polarities cancel. Directly comparing panels B of the two figures, we see that the coronal hole has considerably more positive magnetic flux than the mixed polarity (zero mean-field) quiet Sun.

In panels C, we show the angle to the vertical of the extrapolated linear force-free field \citep[][]{Alissandrakis1981, Gary1989} at 2Mm in the atmosphere, just above the probable formation height of the \ion{Ne}{8} emission \citep[][]{Tu2005a,Tu2005b}. By isolating angles greater than 35 degrees in this region, we attempt to show the locations of ``coronal funnels'' invoked largely to explain the patterning of the \ion{Ne}{8} blue shifts in SUMER rasters \citep[][]{Xia2003, Xia2004, Tu2005a, Tu2005b}. We see that the funnels, more often than not, coalign with the bright \ion{Si}{2} emission at bright supergranular vertices (see, e.g., Figs.~\pref{fig2} and~\pref{fig3}) and overlie magnetic flux regions of the same polarity as the dominant magnetic polarity of the supergranule. Comparing the funnel locations with the \ion{C}{4} blue shift locations (panel E), we see little correspondence at all; the \ion{C}{4} blue shift regions lie immediately outside the funnels. Using the data in panel G of Fig.~\pref{fig8}, we see that, in the quiet Sun, the funnels outline more than 65\% ($\pm$5\% for estimated errors in coalignment with MDI) of the blue shifted material in \ion{Ne}{8} (Fig.~\pref{fig10}). However, in the coronal hole (panel G of Fig.~\pref{fig9}) the correspondence between funnels and blue shifted \ion{Ne}{8} plasma drops dramatically to 35\% ($\pm$5\%; Fig.~\pref{fig11}).

Panels D (of Figs.~\pref{fig8} and~\pref{fig9}) show a simple diagnostic of the degree of magnetic imbalance present, the ``Magnetic Range of Influence'' \citep[MRoI;][]{McIntosh2006}. The MRoI is an estimate of the radial distance needed from a particular magnetic flux concentration to meet enough flux of the opposite polarity to balance and is computed from the full disk (full spatial resolution) MDI magnetogram. The MRoI can be thought of as a crude measure of how open (large MRoI) or closed (small MRoI) the region is. In the quiet Sun, the MRoI map shows few locations greater than 100Mm; these, however, coalign well with the vertex or coronal funnel locations and thus the regions of strong \ion{Ne}{8} blue shift. Conversely, in the coronal hole, we see that there are few locations where the MRoI map drops below 100Mm, and the regions of large MRoI appear to correspond with the global structure of the \ion{Ne}{8} blue shift pattern \citep[Fig. 3 of ][]{McIntosh2006}. Similarly, the vast majority of \ion{C}{4} blue shifts occur in regions where the MRoI is greater than 100Mm and there is net imbalance in the magnetic field. We speculate that the existence of large scale regions of unbalanced magnetic fields in the coronal hole, and the energy that they contain, is important in setting the amount of energy released and hence the magnitude of the blue shifts observed. 

\subsection{Key Points in the Analysis}
Before proceeding we would like to summarize the key points of the analysis presented above:
\begin{description}
\item[\--]{In the coronal hole, the \ion{C}{4} and \ion{Ne}{8} supergranular interior and boundary emission drop by $\sim$40\% compared to their values in the quiet Sun. Further, there is a strong correlation between blue shifts in \ion{C}{4} and \ion{Ne}{8}, and regions of lowest emission. Both are well observed correlations \citep[e.g.,][]{Dere1989, Wilhelm2002, Aiouaz2005}.}
\item[\--]{In the quiet Sun, 65\% of the \ion{Ne}{8} blue shifted plasma lies in regions of unbalanced magnetic flux at bright supergranular vertices, so-called ``coronal funnels'' \citep[e.g.,][]{Xia2003, Xia2004, Tu2005a, Tu2005b}.}
\item[\--]{In the coronal hole, only 35\% of the \ion{Ne}{8} blue shifted plasma lies at bright supergranular vertices. The remaining 65\% of the \ion{Ne}{8} blue shifted plasma lies in regions of unbalanced magnetic flux away from bright network vertices. Further, we agree with the analysis of \citet{Aiouaz2005} that the additional \ion{Ne}{8} blue shift regions are not directly correlated with the observed network structure, contrary to the earlier analysis of \citet{Hassler+others1999}.}
\item[\--]{There is a larger fraction of \ion{C}{4} blue shifting pixels (21\%) in the coronal hole compared to the quiet Sun (4\%). The vast majority of \ion{C}{4} blue shifts (in both cases) occur in regions of unbalanced magnetic flux. This dramatic change in the fractional coverage  of \ion{C}{4} blue shifts matches a suite of sounding rocket \ion{C}{4} observations \citep[e.g.,][]{Dere1989}.}
\item[\--]{There are very few locations of \ion{C}{4} blue shift on supergranular boundaries. By far, the higher percentage (96\%) are located to the interior side of supergranular boundaries; less than 10\% of the \ion{C}{4} blue shifts are within 2Mm of the boundary. In the quiet Sun, the \ion{C}{4} blue shifts neighbor bright supergranular vertices.}
\item[\--]{Some 92\% of coronal hole \ion{C}{4} blue shifts underlie \ion{Ne}{8} blue shifts of comparable, or larger, amplitude. This is compared to only 5\% outside in the quiet Sun. This implies a physical (magnetic) connection between the blue shifted 100,000K and 600,000K plasmas in the transition region. These connected blue shift regions almost uniquely occur in regions of significant magnetic unbalance.}
\end{description}

\section{Discussion}
We have used SUMER raster observations of an equatorial coronal hole and a quiet Sun region to investigate the relationships between the multithermal emitted spectral line intensities and Doppler velocities, the supergranular patterning of the chromospheric network, and the magnetic environment of the plasma. The pictorial and statistical relationships developed for the transition region emission lines of \ion{Si}{2}, \ion{C}{4} and \ion{Ne}{8} suggest that the key to understanding the physical mechanism responsible for coronal heating, as well as the acceleration of the solar wind rests on the correct physical interpretation of the Doppler velocity patterning present in the \ion{C}{4} and, in particular, the relationship of the strong \ion{C}{4} blue shifts regions with the supergranular boundaries and magnetic polarity balance on supergranular spatial scales. While most of the previous analyses \citep[e.g.,][]{Wilhelm2002, Xia2003, Xia2004, Tu2005b}, have presented the emission and Doppler velocities of \ion{C}{4} pictorially, few have compared the \ion{C}{4} Doppler velocity patterning with that observed in \ion{Ne}{8}. The present analysis complements the past results, but provides a new, self-consistent interpretation of the observations.

\begin{figure}
\centering
\plotone{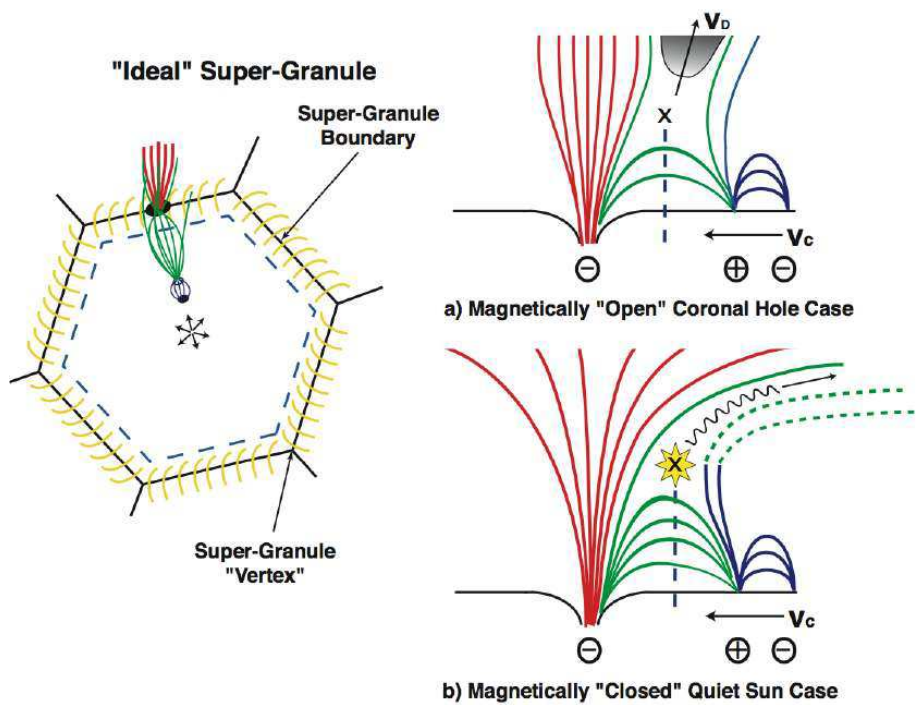}
\caption{Cartoon of a supergranular cell and the action of the magnetoconvection driven magnetic carpet in the magnetically ``open'' network (Coronal Hole) case (a) and the magnetically ``closed'' network (Quiet Sun) case (b). In each case, the network boundary flux is shown in red, while the emerging magnetic dipole is in blue. The green lines of force are those resulting from the reconnection of the emerging advected dipole and the network boundary flux. The X marks the reconnection ÒX-PointÓ where the bulk of the stored energy in the field is released. In case a) with no overlying closed magnetic field, the magnetically released material can ``escape'' to become a tributary to the solar wind while, in case b), with overlying closed quiet Sun the magnetic field releases the bulk of its energy thermally into the loop structure. The blue dashed line indicates the potential locus of \ion{C}{4} blue shift. These cartoons are adapted from those published by \citet{Wang1998} and \citet{Priest2002}. \label{fig12}}
\end{figure}

Figure~\pref{fig12} introduces three simple cartoons in an effort to explain the connection between the  magnetic carpet and the multithermal spectroscopic patterning observed by SUMER in the quiet Sun and in coronal holes. These cartoons are derived from an already published, idealized model of a supergranular cells \citep[][]{Wang1998} and cross-sections \citep[][]{Priest2002} of the cell that are representative of the global magnetic environment in a coronal hole (a) and the quiet Sun (b). 

The upper cartoon (a) illustrates the likely situation in a coronal hole. The small (spatial) scale flux elements anchored in the supergranular boundary (shown in red) are effectively open to interplanetary space because they have the same magnetic polarity as the bulk of the coronal hole. As the small, recently emerged magnetic dipole (shown in blue) is advected to the boundary, the leading polarity of the flux begins to reconnect with the anchored element, creating a new magnetic topology in the supergranular interior (shown in green). A portion of the energy released by the reconnection quickly begins to evaporate cool chromospheric material into the topology \citep[see, e.g.,][]{Yokoyama1998, Czaykowska1999}, where the bulk of the remaining energy is released in the form of kinetic energy. The established flow results in the observed correlation of blue shift in \ion{C}{4} and \ion{Ne}{8} above, and to the supergranular interior side of the topological X-point. The blue dashed lines in the cartoons demonstrate the places in this ideal model where the \ion{C}{4} blue shift regions might occur and qualitatively agrees with the determination from the actual data. The newly created flux topology in the coronal hole supergranular interior will then tend to expand to the cell interior, where the net atmospheric pressure is lower, consistent with the predictions of \citet{Schrijver2003}. We predict that the actual location, lifetime, and strength of the blue shift in the coronal hole will depend on the magnitude of the dipole being advected, its size relative to the field on the boundary and the net imbalance of the field in the cell. This cannot be tested with the observations at hand, but we note that the qualitative dependence of the \ion{Ne}{8} Doppler pattern and the MRoI map add weight to this interpretation. 

In the quiet Sun (cartoon b), we assume that the supergranular-boundary-anchored magnetic flux (red) closes with an opposite polarity piece of magnetic flux on a nearby supergranule (Fig.~\pref{fig3}), such that there is very little probability that any arcade created in the supergranular interior can open into interplanetary space \citep{McIntosh2006}. The convection-driven reconnection of the emerging magnetic dipole creates a new magnetic topology (green) in the supergranular interior that is bound below the pre-existing inter-cell magnetic loop arcade. We anticipate that, as the dipole is advected to the boundary and the reconnection progresses, the bulk of the released energy must result in the evaporative mass-loading of the created arcade and thermal heating of the plasma contained in it, since little of it can be rapidly converted into plasma outflow. In this case, we would expect to see very few \ion{C}{4} blue shifts. We presume that the reconnection sites may only be visible in locations near bright supergranular vertices, i.e., where the magnetic field is locally dominated by one polarity (bright network vertices) and so is effectively radial (or open) \citep[e.g.,][]{Lites2002}. The thermal heating and apparent invisibility of the reconnection X-point in the newly created arcade can qualitatively explain the observed increase in emission and resulting increase in contrast of the supergranular boundaries visible in the quiet Sun. 

We can probabilistically explain the dearth of \ion{C}{4} blue shift regions in the quiet Sun compared to the coronal hole. In the coronal hole, there is net imbalance in the magnetic field and the probability of dipole annihilation increases with the degree of magnetic imbalance. As such, the dipole can be annihilated anywhere in the coronal hole cell, but it has a significantly higher probability of destruction closer to the boundary (where the net magnetic flux is larger). In the quiet-Sun, where the supergranular cell has a near-zero mean-field, the probability of immediately annihilating the advecting dipole is almost negligible, hence the dipoles will live much longer and will probably have to migrate significantly closer to the supergranular boundaries (or vertices) before they meet enough imbalance in the magnetic field to be destroyed. Further, we believe that the relentless magnetoconvection-driven reconnection could explain the prevalent transition region red shift \citep[e.g.,][]{Warren+others1997}. We suspect that the same process of radiative cooling of downflowing material \citep{Muller2005}, observed as ``coronal rain'' \citep[e.g.,][]{Foukal1976, Foukal1977, Ionson1978, Bruner1979, Athay1980} in the vicinity of sunspots over the limb is largely responsible for the differences in the coronal hole and quiet Sun \ion{C}{4} red shifts (Figs.~3, 4 \& 13). Consider the fact that, in the quiet Sun, it appears that much less of the material ejected into the magnetic topology in the upper atmosphere can ``escape'' and therefore, under the influence of gravity it will fall, creating a ``coronal drizzle'' in the supergranular cell. This effect will be less prevalent in the magnetically open coronal hole {\em if} the ejecta form the basis of the solar wind and do not return. We also should note that it is entirely possible that acoustic waves are also present in the down-flowing material and contribute to the net red shift observed \citep[][]{Hansteen1993}. While we cannot tell using the present observations, recent research has demonstrated the connection of transient, propagating low-frequency ($<$5mHz) magneto-acoustic waves in spicules \citep{DePontieu2004} and around supergranular cell boundaries \citep[][]{Jefferies2006} may form a ready source of these waves. 

While we have largely chosen to neglect the discussion of \ion{Ne}{8} blue shifts at bright supergranular vertices, they form an important component of the complete physical description of the observations that we present. The occurrence of coronal magnetic funnels is clearly demonstrated, having impact primarily on the quiet Sun (e.g., Fig.~\pref{fig8}). The very close proximity of the \ion{C}{4} blue shift regions in the quiet Sun to the magnetic funnels suggests that the mechanism invoked to supply mass and energy to the funnel \citep[e.g.,][]{Xia2003} is correct and akin to the magnetic carpet driven ``magnetic exchange reconnection'' model of \citet{Wang1998}, which is thought to load and accelerate mass in a ``polar plume''. If we can assume that the \ion{Ne}{8} blue shifts observed at bright supergranular vertices are near cousins of polar plumes, the additional magnetic flux concentrations on and around the supergranular boundaries in the coronal hole are of significantly smaller spatial scale \citep[$\sim$1000 times smaller at a fraction of an arc-second in radius;][]{Berger2001} and as such one million times smaller in area - a spatial distribution of micro-plumes (``$\mu$-plumes''). Although it is clearly beyond the scope of this article, we draw the reader's attention to the spatial distribution of the spectroscopic characteristics of these ejecta and similarities with other solar phenomena closely tied to supergranular structure: ``spicules'' \citep[e.g.,][]{Secchi1877, Roberts1945, Beckers1968, Sterling2000, DePontieu2004}, bright and dark H-$\alpha$ ``mottles'' \citep[e.g.,][]{AthayThomas1961,Bray1973, BrayLoughead1973}, ``blinkers'' \citep[e.g.,][]{Harrison1997} and ``explosive events'' \citep[e.g.,][]{Innes1997, Bewsher2005}. We are convinced that these phenomena are united by one underlying physical mechanism modified in spectral output by the topology of the magnetic field on the supergranular and global scales \citep[e.g.,][]{Madjarska2006, McIntosh2006c}.

\section{Coupling to the Global Scale of the Solar Atmosphere}
In the previous sections, we have focussed on small-scale ($\sim$20Mm) energy release into the upper solar atmosphere. \citet{McIntosh2006} discussed the influence that the magnetic environment has on the \ion{Ne}{8} emission and Doppler velocities. It was demonstrated that the topological nature (open or closed) and net balance of the magnetic field coupled to the emission and velocities observed for the equatorial coronal hole that was studied with SUMER from 1999 November 3\--8. 

From Fig.~3 of \citet{McIntosh2006}, we see that in the open (large MRoI; net positive field \brr $\ge$ 5G) regions a strong \ion{Ne}{8} blue shift was evident which appeared to get stronger as \brr{} increased. Conversely, in places of (zero mean) closed magnetic field quiet Sun (small MRoI; net magnetic balance \brr $\approx$ 0G) the blue shift pattern is replaced by one that shows a net red shift ($\sim$1km/s), the \ion{Ne}{8} blue shifts are highly localized (into ``funnels'') and the emission increases visibly by a factor of two. 

\begin{figure}
\centering
\plotone{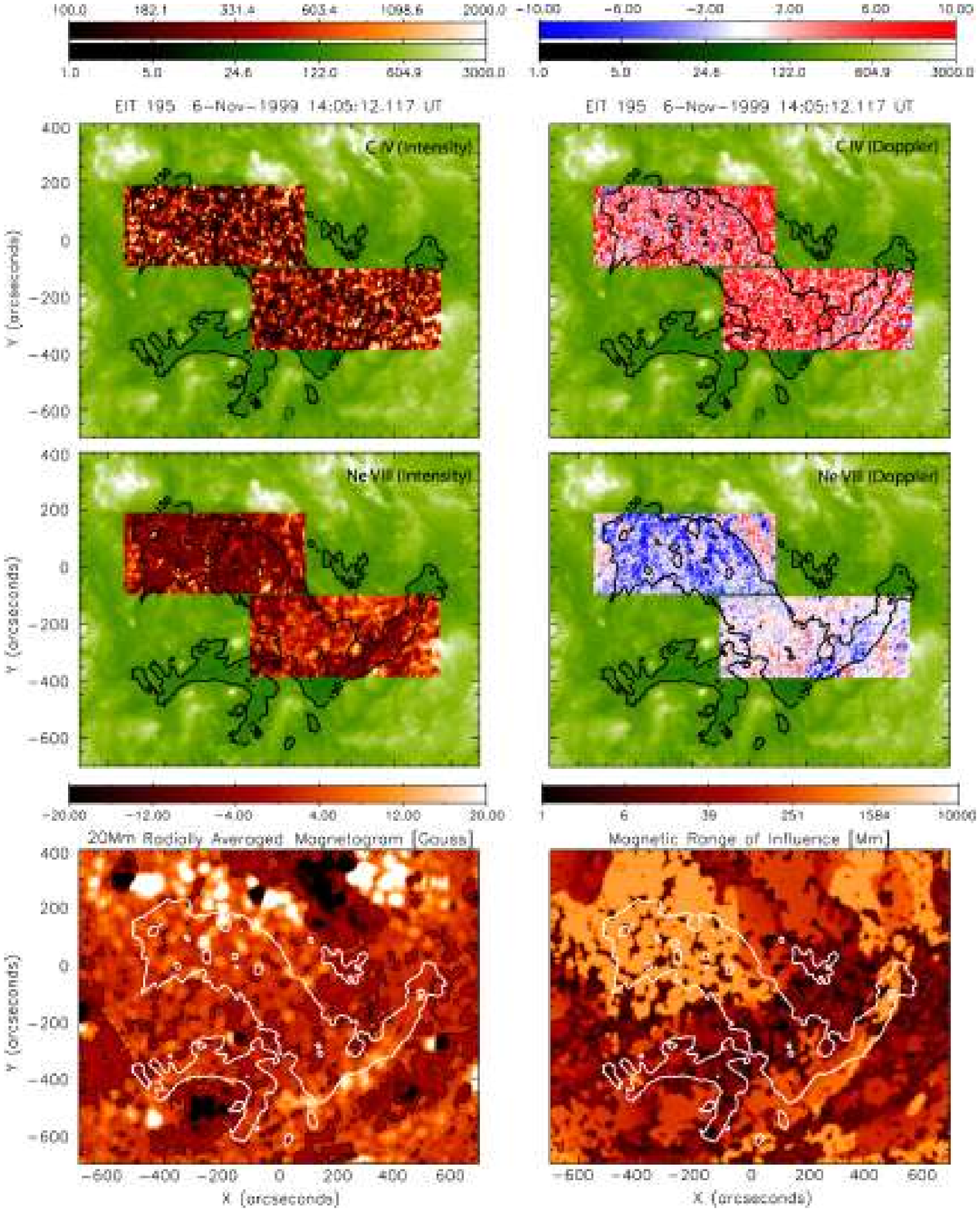}
\caption{An adaptation of Fig.~3 of \citet{McIntosh2006}. The top two rows show the SUMER \ion{C}{4} 1548~\AA{} and \ion{Ne}{8} 770~\AA{} integrated line intensities (Units: Counts) and Doppler-velocities (km~s$^{-1}$), respectively. The bottom row shows the supergranular radially averaged magnetic field strength (\brr; Gauss) and the ``Magnetic Range of Influence'' (MRoI; Mm). On each of the panels in the figure, we show the EIT 195~\AA{} 150~DN contour to outline the coronal hole region. Additionally, in the lower left panel, we show the thin black contour that designated the magnetic ``neutral line'', where \brr = 0~G. \label{fig13}}
\end{figure}

Figure~\pref{fig13} is a simple adaptation of Fig.~3 of \citet{McIntosh2006}, where we now include the global dependence of the \ion{C}{4} emission and Doppler velocity structures in the equatorial coronal hole (top row) with those of the hotter \ion{Ne}{8} (middle row) and the two magnetic field balance diagnostics discussed above (bottom row). The spatial correspondence between significant \ion{Ne}{8} blue shifts and regions of unbalance in the photospheric magnetic field is clear. Similarly, one can observe the increased coverage (4\% increasing to 21\%) of significant \ion{C}{4} blue shift in the same places. \citet{McIntosh2006} demonstrate that the \ion{Ne}{8} Doppler velocity partitioning of the coronal hole in the central region (x=-100\arcsec{},200\arcsec{}, y=-300\arcsec{}, -100\arcsec{} \-- seen most clearly in the \ion{Ne}{8} intensity maps) is due to the local closure of the magnetic fields (low values of MRoI and nearly balanced magnetic fields); this region is consistent with quiet Sun energy release (enhanced \ion{Ne}{8} emission, isolation of blue shifted material to bright network vertices and few \ion{C}{4} blue shift locations). Conversely, we also see that the the open field regions determined by the magnetic diagnostics (high MRoI and unbalanced magnetic fields) point to regions that show a marked increase in the number of pixels showing significant blue shifting plasma in \ion{C}{4} which, as we have shown, invariably lie below the blue shifting \ion{Ne}{8} regions. We also point the reader to the spatial dependence of the \ion{C}{4} Doppler velocities outside the EIT 150~DN contour and in the central (\ion{Ne}{8} intensity enhanced) region of the coronal hole. There is a clear increase in the magnitude of the red shift ($\sim$7km/s) and lack of the cellular contrast (the red shift appears to fill the entire cell) that is very visible in the magnetically unbalanced portions of the coronal hole (e.g., Figs.~\pref{fig6} and~\pref{fig8}).

We deduce that the global aspects of the spectroscopic signals observed (thermally) through the plasma, when compared to the magnetic flux balance, are consistent with the presence of magnetoconvection-driven reconnection as the dominant supply of energy to the plasma.

Unfortunately, the time taken to produce the two SUMER equatorial coronal hole rasters shown here ($\sim$20 hours) does not allow us to discuss the spectroscopic evolution of the transition region plasma and place it in the context of the EUV corona, and the location where the EUV corona and transition region plasmas appear to be thermally partitioned, or ``nested'', in the central portion of the coronal hole. We know from the movie accompanying \citet{McIntosh2006} that the EUV coronal hole does not appear to evolve dramatically over the time taken to compose the rasters shown. \citet{McIntosh2006} briefly discuss the ``nesting'' of the coronal hole and propose that there are either two distinct coronal holes present (one to the North and one to the South with the increased emission region in the center underlying the divergent magnetic topologies of each) or that there is a very different loading and heating of the EUV plasma above the same (apparently closed) region in the transition region. While the former is the simplest explanation, we now discuss the latter. 

It is possible that the magnetic environment of the EUV coronal hole is isolated from the quiet Sun and forms a perimeter isolating the region where the MRoI drops dramatically, the emission increases and the two velocity patterns (\ion{C}{4} and \ion{Ne}{8}) adopt their quiet Sun form. In this case, we must explain why the column depth of the plasma observed in the 195\AA{} EIT passband is (almost) uniformly reduced above this region, as is the Kitt Peak Vacuum Telescope \ion{He}{1} 10830\AA{} spectroheliogram equivalent width\footnote{Similarly, anomalous changes in other chromospheric line widths observed when crossing from quiet-Sun to plage regions (regions of net-zero mean field flux to ones with very large polarity imbalance) that have been largely attributed to changes in the ``micro-turbulence'' \citep[][]{Simon1980}, might be considered simply as the natural interface between a thermal (quiet-Sun; mean-zero field) and a non-thermal (plage; closed topology but large non-zero mean field) process. The work of \citet[][]{Worrall1972, Worrall1973, vanBreda1995, Worrall2002} on other chromospheric spectral abnormalities may be considered with the same perspective.}. The apparent mass reduction in the region (195\AA; reduced column mass) and presence of excess non-thermal processes (10830\AA; reduction of line widths) are both consistent as coronal hole ``markers'' in the sense of the simple energy partitioning by the magnetic topology (open \-- kinetic; closed \-- thermal). If the closed field region is imbedded in a globally open magnetic topology, it may simply be that filling the coronal volume in that region is only possible from the quiet-Sun coronal funnels at the bright network vertices where we see \ion{Ne}{8} blue shifts. From limited sources of mass, that would be difficult, and so the dispersal of mass must affect both the 195\AA{} and 10830\AA{} measurements in the manner observed. It was noted in \citet{McIntosh2006} (and independently \-- J.~B. Gurman 2005, private communication) that there is a slight increase ($\sim$10DN) in 195\AA{} intensity overlying this closed region, but it had very low contrast with the remainder of the coronal hole. Clearly, we are at an impasse; both of these situations are possible (as are several more), but they may only be resolved by detailed magnetic extrapolation or by watching the evolution of the coronal hole for several more rotations before and after the region was observed with SUMER. We leave this investigation for future work, but note that the global spectroscopic structure of the coronal hole did not vary over the 5 days that it was observed \citep[][]{Davey2006}.

\subsection{Influence on the Heliosphere}
Many other markers of the coupling between the energy equation (initial energy release and distribution) and the magnetic flux balance of the solar plasma may have already been widely observed, but not linked. These would appear to have an effect on the heliospheric system as a whole, but are too numerous to discuss in this paper, instead we cite a few. Systematic \ion{He}{1} 10830\AA{} line asymmetries observed at the solar South pole near solar minimum (1995 October 17) have been attributed to a strong outflow ($\sim$8km/s) of chromospheric material \citep[][]{Dupree1996} and occur at a time when there is a relatively strong unbalanced field at the pole. The magnetic flux that diffuses to the polar regions over the course of the solar cycle creates a long-lived ``polar crown'' coronal hole at solar minimum. This excessive imbalance in the polar magnetic field, we suspect, will act as a reservoir of kinetic energy for the very high speed winds observed at high heliospheric latitudes at the same phase of the solar cycle \citep{McComas1998} while the evolving mixture of open and closed regions and the resulting energy partitioning of the plasma explain the complex heliospheric structure observed at other phases of the solar cycle \citep[e.g.,][]{Geiss1995, McComas2000, Smith2003}. One can wonder about the possible implications of this result for other observations that appear to connect coronal holes, chromospheric material and the solar wind \citep[e.g.,][]{McIntoshLeamon2005, Leamon2006}.

The role of the global magnetic topology as a means of the controlling how the energy released into the plasma is used may have implications for the ``FIP effect'' \citep[e.g.,][]{Raymond1999, Feldman2000, Laming2006, Young2006, Ko2006}. The primary signature of the FIP effect is that elements with a low first ionization potential (FIP; e.g., Na, Si, Al, Ca, Fe and Ni) are often enhanced over those with a high FIP (e.g., He, N, O, Ne and Ar) in the slow solar wind by a factor of several while the fast solar wind shows little or no elemental fractionation at all. We speculate that the contrast in the FIP-related enhancement of certain atomic species is tied to how and where the mass and energy are delivered to and used by the plasma. We have seen that the fast solar wind generally originates from open magnetic regions of sizable magnetic imbalance and that the energy is delivered in a mostly non-thermal (kinetic) form with mass that originates in the well mixed chromosphere. Conversely, in the case of the slow solar wind, originating from the quiet Sun and a largely closed magnetic topology permeated by small ``coronal funnels'', will ensure a complex, mostly thermal, energy delivery to the plasma. It is possible that enhanced fractionation can take place in the plasma it is transported through the magnetic topology and eventually evaporated out of the corona and into the heliosphere in a fashion that depends on the exact magnetic topology \citep[e.g.,][and references therein]{Schwadron2003}. While a full investigation into the connection between the result presented in this paper and the heliospheric plasma is beyond the scope of this paper we can see that there is significant potential.

\section{Conclusion}
Together, the \soho{} (SUMER, EIT and MDI) observations, statistical relationships and cartoon representations lead us to the conclusion that the observed Doppler velocity and emission patterning of the upper transition region and low solar corona is consistent with the action of convection-driven magnetic field emergence and reconnection: the magnetic carpet \citep[][]{Schrijver1997, Schrijver1998, PriestSchrijver1999, Priest2002, PriestForbes2000, Wang1998}.

We have demonstrated that the driven reconnection events largely neighbor the supergranular boundaries and propose that they contain plasma that is loaded and driven by the magnetic carpet's relentless stirring and destruction of the injected and advected magnetic flux \citep[][]{Schrijver2003}. We have deduced that while the net magnetic flux on the scale of a supergranule controls the injection rate of mass and energy into the transition region plasma it is the global magnetic topology of the plasma that dictates whether the released ejecta provides thermal input to the quiet solar corona or becomes a tributary to the solar wind.

The magnetoconvection-driven reconnection and resulting ejecta that we observe and have discussed has significant impact on the energetics of the outer solar atmosphere. The same must be true for stellar atmospheres for which dynamo-driven, cyclic magnetic field behavior has been observed in their optical/UV/EUV radiative output. The hypothesis presented in this Paper can be directly tested by the instruments on the upcoming Solar-B satellite and should provide motivating science for the Solar Probe and Solar Orbiter Missions currently in the pre-proposal phase.

\acknowledgements 
We would like to thank Drs. David Alexander, Tom Bogdan, Robin Canup, Joe Gurman, Stuart Jefferies, Philip Judge, Bob Leamon, Nathan Schwadron, Meredith Wills-Davey and several anonymous referees for kind assistance, helpful discussions and comments on the manuscript that have greatly influenced the ideas presented. This material is based upon work carried out at the Southwest Research Institute that is supported in part by the National Aeronautics and Space Administration under grants issued under the Living with a Star, Sun-Earth Connection Guest Investigator Programs and Solar Data Analysis Center, specifically Grants NAG5-13450 and NAG5-11594 (to DMH) in the early and NNG05GM75G, NNG06GC89G, NNG05GQ70G (to SWM) in the closing phase of the work reported. The SUMER project is financially supported by DLR, CNES, NASA and the ESA PRODEX Program (Swiss contribution). SUMER is part of \soho, the Solar and Heliospheric Observatory, of ESA and NASA. This paper is dedicated to the memory of Alan Stuart McIntosh (1988 - 1994).

\clearpage

\end{document}